\documentclass[epj-spec]{svjour}
\usepackage{epsfig}
\begin{document}
\title{The total charm cross section}
\author{R. Vogt\inst{1}\inst{2}
\thanks{\email{vogt2@llnl.gov}} 
}
\institute{Lawrence Livermore National Laboratory, Livermore, CA, USA 
\and Department of Physics, UC Davis, Davis, CA, USA}
%
\abstract{
We assess the theoretical uncertainties on the total charm cross
section.  We discuss the importance of the quark mass, the scale choice
and the parton densities on the estimate of the uncertainty.  
We conclude that due to the small charm quark mass, which amplifies the
effect of varying the other parameters in the calculation, the uncertainty on 
the total charm cross section is difficult to quantify.
} 
\maketitle
\section{Introduction}
\label{intro}

Open charm measurements date back to the late 1970s when $D$ 
and $\overline D$ mesons were first detected, completing the
picture of the fourth quark begun when the $J/\psi$ was detected in $p$Be
and $e^+ e^-$ interactions.  The charm quark was postulated to have a mass
between 1.2 and 1.8 GeV, within the regime of perturbative quantum 
chromodynamics (pQCD).  Because of its rather large mass relative to the
$u$, $d$ and $s$ quarks, it is 
possible to calculate a total $c \overline c$ cross section, not
the case for lighter flavors.  Charm hadrons are usually
detected two ways.  The reconstruction of decays to charged hadrons 
such as $D^0 \rightarrow K^- \pi^+$ (3.8\%) and $D^+ \rightarrow K^- \pi^+ 
\pi^+$ (9.1\%) gives the full momentum of the initial $D$ meson, yielding
the best direct measurement.  Charm can also be detected indirectly via
semi-leptonic decays to leptons
such as $D \rightarrow K l \nu_l$ although
the momentum of the parent $D$ meson remains unknown.  Early measurements
of prompt leptons in beam dump experiments assumed that the density of the dump
was high enough to absorb semi-leptonic decays of non-charm hadrons,
leaving only the charm component.  At modern colliders, it is not possible to
use beam dumps to measure charm from leptons but, at sufficiently
high $p_T$, electrons from charm emerge from hadronic cocktails 
\cite{star,phenix}.

Although $D$ mesons alone are often used to calculate the
total $c \overline c$ cross section, other charm hadrons also exist.
The excited $D$ states, $D^*$s, decay primarily to charged and neutral
$D$ mesons.  The charm-strange meson, the $D_s$, decays to charged hadrons
as well as semi-leptonically.  The lowest mass charm baryon
is the $\Lambda_c^+$ which decays primarily to $\Lambda(uds)$ 
but also decays to $pK^- \pi^+$ (2.8\%) and semi-leptonically with a 4.5\% 
branching ratio.  The heavier ground
state charm baryons and their excited states ($\Sigma_c$ and higher) decay
via the $\Lambda_c$s.  The charm-strange baryons are assumed to be 
a negligible contribution to the total cross section.

Extracting the total charm cross section from data
is a non-trivial task.  To go from
a finite number of measured $D$ mesons in a particular decay channel
to the total $c \overline c$ cross section one must: divide by the branching 
ratio for that channel; correct for the luminosity, $\sigma_D = N_D/ 
{\cal L}t$; extrapolate to full phase space from the finite detector 
acceptance; divide by two to get the pair cross section from the single $D$s;
and multiply by a correction factor \cite{Mangano} to account for the 
unmeasured charm hadrons.  There are assumptions all along the way.  The most
important is the extrapolation to full phase space.  Before QCD calculations
were available, the data were extrapolated assuming
a power law for the $x_F$ distribution, related to the longitudinal
momentum of the charm hadron by $x_F = p_z/(\sqrt{S}/2) = 2 m_T 
\sinh y/\sqrt{S}$.  The canonical parametrization $(1 - x_F)^c$ 
was used where $c$
was either fit to data over a finite $x_F$ range or simply assumed.  These
parametrizations led to large overestimates of the total cross section
when $0<c<2$ was assumed, especially when data were taken only near
$x_F = 0$.  Lepton measurements were more conservative but were typically at
more forward $x_F$.

Rather than assess the uncertainties in the data, here 
we address the theoretical
uncertainties in the calculation of the charm and bottom cross sections.
Since the data are generally taken in a finite kinematic region, we begin with 
the calculation of the inclusive distributions to the Fixed-Order 
Next-to-Leading Logarithm (FONLL) level and then discuss the
total cross section calculations to next-to-leading order (NLO), 
the most accurate calculation of the total cross section over all energies.

We calculate the transverse momentum, $p_T$, distributions of
charm and bottom quarks, the charm and bottom hadron distributions
resulting from fragmentation and,
finally, the electrons produced in semi-leptonic decays of the hadrons
\cite{CNV}.  We then calculate the total charm and bottom cross sections, both
by the integral over the inclusive $p_T$ distribution and by integrating the
total partonic cross section.  
At each step, we clarify the theoretical framework as well as the
parameters and phenomenological inputs. 
Our final prediction is thus not a single number but rather an
uncertainty band which has a reasonably large probability of containing
the `true' theoretical prediction. We show that applying this procedure blindly
may lead to an apparent discrepancy in the two methods, particularly for
charm production.  We explain why this seems to be the case as well as
why, when the calculations are done consistently, there is no discrepancy.
The theoretical uncertainties in both methods of obtaining the total cross
section are estimated as extensively as possible.  We show that, for
charm production, the theoretical uncertainty on the total cross section
is difficult to quantify in a reliable way.

\section{Total heavy flavor cross sections from integrated inclusive 
distributions}
\label{fonll-sec}

We first discuss how the total cross section and its accompanying uncertainty
is obtained from inclusive single particle distributions.  We begin with a
description of the calculated single electron spectrum since heavy flavor 
hadrons are often observed through their semi-leptonic decays, particularly
at colliders where direct reconstruction of heavy flavored hadrons at low
$p_T$ is difficult.  Reconstructed $D$ and $B$ meson decays can only be
used to obtain the total heavy flavor cross section if they are measured down
to $p_T = 0$, difficult at colliders.  Since the RHIC experiments are 
designed to measure low $p_T$ hadrons, the full $p_T$ distribution can be
accessed.  Thus STAR has reconstructed $D^0
\rightarrow K^\pm \pi^\mp$ decays to $p_T \sim 0$ in addition to
their single electron measurement \cite{star}.  PHENIX has measured the
single electron spectra from heavy flavor decays alone \cite{phenix}.

The theoretical prediction of the electron spectrum includes 
three  main components: the $p_T$ and rapidity
distributions of the heavy quark, $Q$, in $pp$ collisions at $\sqrt{S} =
200$~GeV, calculated in perturbative QCD; fragmentation of the
heavy quarks into heavy hadrons, $H_Q$, described by
phenomenological input extracted from $e^+e^-$ data; and the decay of
$H_Q$ into electrons according to spectra available from other 
measurements. This
cross section is schematically written as
\begin{eqnarray}
\frac{E d^3\sigma_e}{dp^3} = \frac{E_Q d^3\sigma_Q}{dp^3_Q} \otimes 
D(Q\to H_Q) \otimes f(H_Q \to e) \; , 
\label{siginclusive}
\end{eqnarray}
where $\otimes$ denotes a generic convolution. The electron decay
spectrum term, $f(H_Q \to e)$, also implicitly accounts for the proper 
branching ratio to leptons.

The distribution $E d^3\sigma_Q/dp^3_Q$ is evaluated at the FONLL 
level, implemented in Ref.~\cite{Cacciari:1998it}.
In addition to including the full fixed-order NLO
result~\cite{Nason:1987xz,Beenakker:1990ma}, the FONLL calculation also
resums~\cite{Cacciari:1993mq} large perturbative terms proportional to
$\alpha_s^n\log^k(p_T/m)$ to all orders with next-to-leading logarithmic (NLL)
accuracy (i.e. $k=n,\,n-1$) where $m$ is the heavy quark mass. 

The perturbative
parameters are the heavy quark mass and the value of the strong coupling,
$\alpha_s$, while the parton densities are a nonperturbative input. 
We take central values of 1.5~GeV for charm and 4.75~GeV for bottom
and vary the masses between 1.3 and 1.7~GeV for charm and 
4.5 and 5~GeV for bottom to estimate the resulting mass uncertainties.  

Since the
FONLL calculation treats the heavy quark as an active light flavor at $p_T >>
m$, the number of light flavors used to calculate $\alpha_s$ includes the
heavy quark, i.e. $n_{\rm lf} + 1$ where, for charm, $n_{\rm lf} = 3$ ($u$, $d$
and $s$).  The same number of flavors, $n_{\rm lf} + 1$, is also used in the 
fixed-order scheme where the quark mass is finite.  However, in other 
fixed-order calculations, {\it e.g.} to leading and next-to-leading order,
the number of light flavors is fixed to $n_{\rm lf}$.
The QCD scale at five flavors, $\Lambda^{(5)}$, is set to 0.226 GeV, as in the
CTEQ6M parton densities \cite{cteq6m}. 

The perturbative calculation also depends on the
unphysical factorization ($\mu_F$) and renormalization ($\mu_R$) scales.  The
sensitivity of the cross section to their variation can be used to estimate the
perturbative uncertainty due to the absence of higher orders. We have taken
$\mu_{R,F} = \mu_0 = \sqrt{p_T^2 + m^2}$ as a central value in the inclusive
distributions and varied
the two scales independently within a  `fiducial' region defined by  $\mu_{R,F}
= \xi_{R,F}\mu_0$ with $0.5 \le \xi_{R,F} \le 2$ and $0.5 \le  \xi_R/\xi_F \le
2$. In practice, we use the following seven sets: $\{(\xi_R,\xi_F)\}$ =
\{(1,1),  (2,2), (0.5,0.5), (1,0.5), (2,1), (0.5,1), (1,2)\}.  The
uncertainties stemming from mass and scale variations are added in quadrature.
The envelope containing the resulting curves defines the uncertainty. 

The fragmentation functions, $D(c\to D)$ and $D(b\to B)$, where $D$ and $B$
indicate a generic admixture of charm and bottom hadrons, are consistently
extracted from $e^+e^-$ data in the context of FONLL~\cite{Cacciari:2002pa}.
Using the Peterson {\it et  al.} fragmentation
function~\cite{Peterson:1982ak}, with standard parameter choices $\epsilon_c
\simeq
0.06 \pm 0.03$ and $\epsilon_b \simeq 0.006 \pm 0.003$, does not provide
a valid description of fragmentation to FONLL.
 
The measured spectra for primary $B\to e$ and $D \to e$ decays are modeled and
assumed to be equal for all bottom and charm hadrons respectively.
The contribution of
electrons from secondary $B$ decays, $B\to D\to e$, was obtained
by convoluting the $D\to e$ spectrum with a parton-model prediction of
$b\to c$ decay.  The resulting electron spectrum is very soft, giving a
negligible contribution to the total.
The decay spectra are normalized using the branching ratios
for bottom and charm hadron mixtures \cite{Eidelman:2004wy}:
BR$(B\to e) = 10.86 \pm 0.35$\%, BR$(D\to e) = 10.3 \pm 1.2$\%,
and BR$(B\to D\to e) = 9.6 \pm 0.6$\%.

\begin{figure}[htbp]
\setlength{\epsfxsize=0.95\textwidth}
\setlength{\epsfysize=0.5\textheight}
\centerline{\epsffile{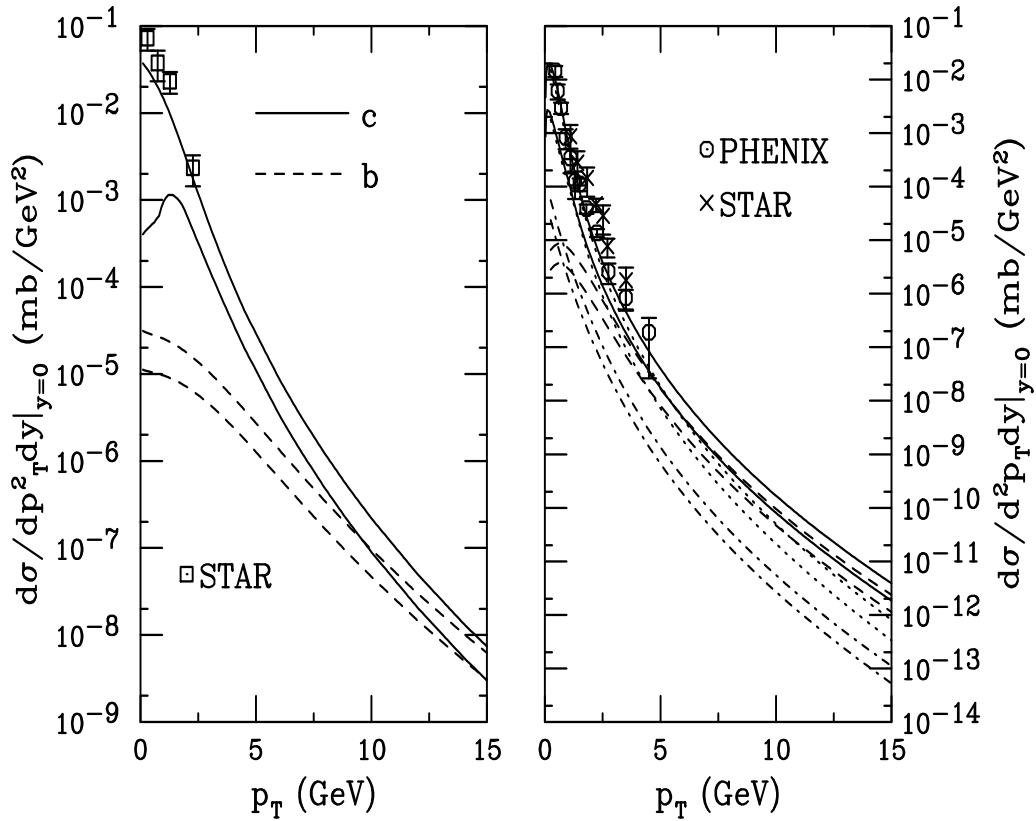}}
\caption[]{Left-hand side: The FONLL $p_T$ bands for $c$ and $b$ quark 
production compared to the STAR $D$ meson data \protect\cite{star}.
Right-hand side: The heavy flavor contributions to the single electron
spectra: $D \rightarrow eX$ (dotted), $B \rightarrow lX$ (dashed), 
$B \rightarrow D \rightarrow lX$ (dot-dashed) and the sum (solid). 
is compared to the PHENIX \protect\cite{phenix} and STAR \protect\cite{star}
data.}
\label{inclusive}
\end{figure}

We first present the transverse momentum
distributions for charm and bottom quarks on the left-hand side of 
Fig.~\ref{inclusive}.  The theoretical uncertainty bands for the
two distributions in Fig.~\ref{inclusive} is obtained by summing 
the mass and scale uncertainties in quadrature so that
\begin{eqnarray}
 \frac{d\sigma_{\rm max}}{dp_T} & = &
\frac{d\sigma_C}{dp_T}
+ \sqrt{\Bigg(\frac{d\sigma_{\mu ,{\rm max}}}{dp_T} -
\frac{d\sigma_C}{dp_T}\Bigg)^2
+ \Bigg(\frac{d\sigma_{m, {\rm max}}}{dp_T} -
\frac{d\sigma_C}{dp_T}\Bigg)^2} \label{fonllmax} \\
\frac{d\sigma_{\rm min}}{dp_T} & = & \frac{d\sigma_C}{dp_T}
- \sqrt{\Bigg(\frac{d\sigma_{\mu ,{\rm min}}}{dp_T}
- \frac{d\sigma_C}{dp_T}\Bigg)^2
+ \Bigg(\frac{d\sigma_{m, {\rm min}}}{dp_T}
- \frac{d\sigma_C}{dp_T}\Bigg)^2} \label{fonllmin} \,\, .
\end{eqnarray}
where $C$ is the distribution for the central value,
$\mu$, max ($\mu$, min) is the maximum (minimum) cross section
obtained by choosing the central value with the scale factors 
in our seven fiducial sets,
and $m$, max ($m$, min) is the maximum (minimum) cross section obtained
with $\xi_R=\xi_F=1$ and the lower and upper limits on the quark mass 
respectively.
There is, however, considerable arbitrariness in the choice of the method used 
to assess the theoretical uncertainties.  In fact, the meaning of the 
theoretical error due to unknown higher order effects is, to a large extent, 
subjective.  The recipe we follow is often used in calculations of cross 
sections at hadron colliders and is similar to the one used to compute heavy 
flavor cross sections at the Tevatron (see 
Refs.~\cite{Cacciari:2003uh,Cacciari:2003zu,Cacciari:2004}).  By experience, 
we assign a probability of 80-90\% that the true result lies within the band.

Note that the charm quark uncertainty band is enlarged at low
$p_T$ due to the large value of $\alpha_s$ at low
scales and the increased sensitivity of the cross section to the
charm quark mass. In Ref.~\cite{CNV}, we also noted that, due to the fairly hard
fragmentation function,  the $D$ meson and $c$ quark
distributions begin to differ outside the uncertainty bands 
only for $p_T > 9$~GeV while the $b$ quark and $B$ meson bands overlap 
over all $p_T$.

The single electron uncertainty bands from $D \rightarrow e$, $B \rightarrow e$
and $B \rightarrow D \rightarrow e$ decays
as well as the sum are compared to the
STAR \cite{star} and PHENIX \cite{phenix} data on the right-hand side of
Fig.~\ref{inclusive}.  As expected, $B \rightarrow D \rightarrow e$
is a negligible contribution to the total.  While $D \rightarrow e$ decays
dominate at low $p_T$, the $B \rightarrow e$ contribution begins to dominate
at higher $p_T$.  The two uncertainty bands cross each other in the region
$3.5 < p_T < 12$ GeV.  The region of crossover is rather broad since we
consider the $c$ and $b$ quark mass and scale uncertainties to be uncorrelated.
If the scale uncertainties were assumed to be correlated, the crossover region
would be narrower, as shown in Ref.~\cite{dgvw}.  However, for a true measure of
the uncertainty, we cannot assume that the scales are correlated.  The
PHENIX measurement is in relatively good agreement with the upper edge of the
uncertainty band in Fig.~\ref{inclusive}
while the STAR data tend to lie a factor of 4-5 above the
central value, falling well above the band.

If the distributions shown here are integrated over all phase space, the 
`perturbative' inputs used in the calculation lead to a FONLL total $c\bar c$ 
cross section in $pp$ collisions of 
\begin{eqnarray}
\sigma_{c\bar c}^{\rm FONLL} = 
256^{+400}_{-146} \, \, \mu{\rm b}
\label{sigccfonll}
\end{eqnarray} 
at $\sqrt{S} = 200$~GeV \cite{CNV}.  
The corresponding NLO  prediction \cite{CNV} is 
\begin{eqnarray}
\sigma_{c\bar c}^{\rm NLO} = 
244^{+381}_{-134} \, \, \mu{\rm b} \, \, .
\label{sigccfonll-nlo}
\end{eqnarray} 
The theoretical uncertainty is evaluated as described above. 
Thus the two calculations are equivalent at the total cross section level 
within the large perturbative uncertainties, as expected.  The total cross
section for bottom production is \cite{CNV} 
\begin{eqnarray}
\sigma_{b\bar b}^{\rm FONLL} = 
1.87^{+0.99}_{-0.67} \, \, \mu{\rm b} \, \, .
\label{sigbbfonll}
\end{eqnarray} 

Because the FONLL
and NLO distributions tend to coincide at small $p_T$ and the
total cross section is dominated by the low $p_T$ region, the total cross 
sections and their uncertainties are nearly equal in the FONLL and NLO 
approaches.  Earlier
papers~\cite{Vogt:2001nh} used $m = 1.2$~GeV and $\mu_R = \mu_F =
2\sqrt{p_T^2 + m^2}$ as reference parameters for charm production. 
With this choice we find
$\sigma_{c\bar c}^{\rm NLO} = 427$~$\mu$b, within the calculated
theoretical uncertainty band.

\section{Total heavy flavor cross section from total partonic cross sections}
\label{nlo-sec}

The total partonic cross section has only been completely calculated to NLO
\cite{Nason:1987xz}.  
Some NNLO calculations are available near threshold, applicable for 
$\sqrt{S} \leq 20-25$ GeV \cite{KLMVcc,KVcc}.
The NLO corrections to the leading order (LO) cross sections are relatively
large, $K_{\rm th} = \sigma_{\rm NLO}/\sigma_{\rm LO} \sim 2-3$, depending on
$\mu$, $m$ and the parton densities \cite{RVkfac}.  The NNLO corrections are
about as large at next-to-next-to-leading logarithm \cite{KLMVcc} but decrease
to less than $K_{\rm th}$ when subleading logs are included \cite{KVcc}. 
Scaling functions \cite{Nason:1987xz} proportional to logs of 
$\mu^2/m^2$ are used to calculate the total
cross section to NLO.

The hadronic cross section in $pp$ collisions can
be written as
\begin{eqnarray}
\sigma_{pp}(S,m^2) & = & \sum_{i,j = q, \overline q, g} 
\int dx_1 \, dx_2 \, 
f_i^p (x_1,\mu_F^2) \,
f_j^p(x_2,\mu_F^2) \, \widehat{\sigma}_{ij}(s,m^2,\mu_F^2,\mu_R^2)
\label{sigpp}
\end{eqnarray}
where $x_1$ and $x_2$ are the fractional momenta carried by the colliding
partons and $f_i^p$ are the proton parton densities.
The partonic cross section is
\begin{eqnarray}
\widehat{\sigma}_{ij}(s,m,\mu_F^2,\mu_R^2) & = & 
\frac{\alpha_s^2(\mu_R^2)}{m^2}
\left\{ f^{(0,0)}_{ij}(\rho) \right. \nonumber \\
 & + & \left. 4\pi \alpha_s(\mu_R^2) \left[f^{(1,0)}_{ij}(\rho) + 
f^{(1,1)}_{ij}(\rho)\ln\bigg(\frac{\mu_F^2}{m^2} \bigg) \right] 
+ {\cal O}(\alpha_s^2) \right\}
\,\, 
\label{sigpart}
\end{eqnarray}
where $\rho = 4m^2/s$ and 
$f_{ij}^{(k,l)}$ are the scaling functions to NLO \cite{Nason:1987xz}.

At small $\rho$, the ${\cal O}(\alpha_s^2)$ and ${\cal O}(\alpha_s^3)$
$q \overline q$ and the ${\cal O}(\alpha_s^2)$ $gg$ scaling functions 
become small while the ${\cal O}(\alpha_s^3)$ $gg$ and $qg$ scaling functions
plateau at finite values.  Thus, at collider energies, the total cross sections
are primarily dependent on the small $x$ parton densities and phase space.

The total cross section does not depend on any kinematic variables, 
only on the quark mass,
$m$, and the renormalization and factorization scales with central
value $\mu_{R,F} =
\mu_0 = m$.  The heavy quark is always considered
massive in the calculation of the total cross section and is thus not an active
flavor in the production calculation.  Therefore, the number 
of light quark flavors, $n_{\rm lf}$, does not include
the heavy quark while the FONLL calculation uses $n_{\rm lf} +1$ flavors since
the heavy quark is an active flavor at high $p_T$, as described in 
Section~\ref{fonll-sec}.  

\begin{figure}[htbp] 
\setlength{\epsfxsize=0.95\textwidth}
\setlength{\epsfysize=0.3\textheight}
\centerline{\epsffile{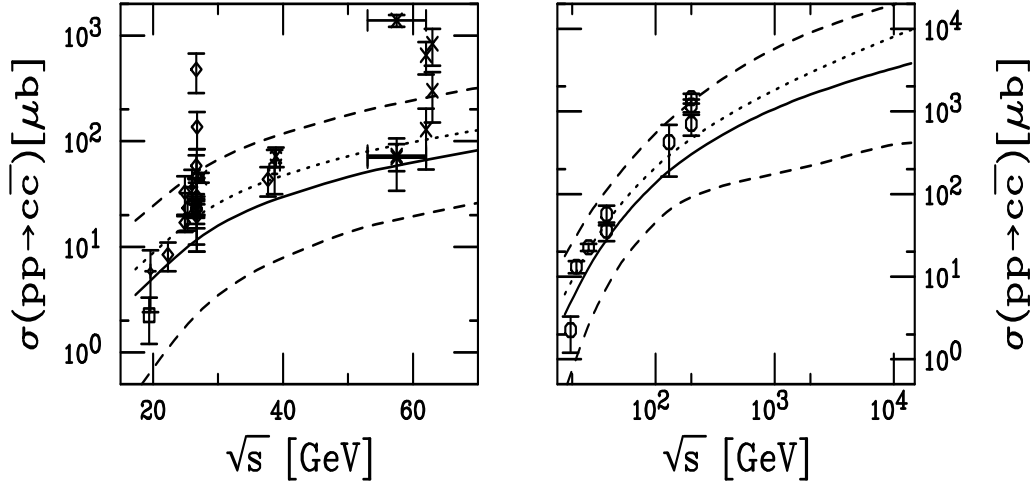}}
\caption[]{The NLO total $c \overline c$ cross sections as a function of
$\sqrt{S}$ for $\sqrt{S} \leq 70$ GeV (left-hand side) and 
up to 14 TeV (right-hand side) calculated with the CTEQ6M parton densities.  
The solid curve is the central result;
the upper and lower dashed curves are the upper and lower edges of the 
uncertainty band.  The dotted curves are calculations with $m = 1.2$ GeV,
$\mu_F = \mu_R = 2m$.
}
\label{totcharm}
\end{figure}

\begin{figure}[htbp] 
\setlength{\epsfxsize=0.95\textwidth}
\setlength{\epsfysize=0.3\textheight}
\centerline{\epsffile{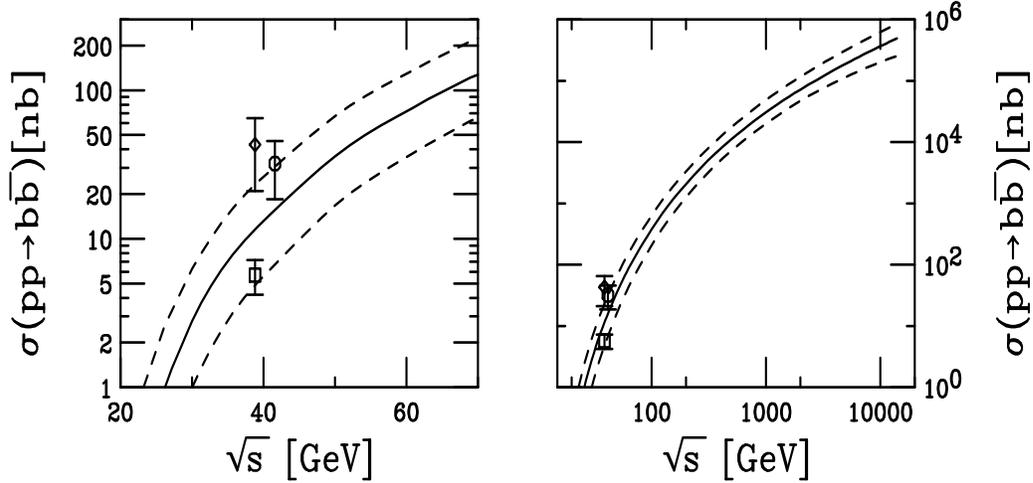}}
\caption[]{The NLO total $b \overline b$ cross sections as a function of
$\sqrt{S}$ for $\sqrt{S} \leq 70$ GeV (left-hand side) and
up to 14 TeV (right-hand side) calculated with the CTEQ6M parton densities.  
The solid curve is the central result;
the upper and lower dashed curves are the upper and lower edges of the 
uncertainty band.  
}
\label{totbottom}
\end{figure}

The theoretical uncertainty on the total cross
section is studied within the same fiducial region as the $p_T$ 
distributions with the upper and lower limits of the uncertainty band determined
as in Eqs.~(\ref{fonllmax}) and (\ref{fonllmin}).  The energy dependence of
the charm and bottom total cross sections is shown in Figs.~\ref{totcharm}
and \ref{totbottom} respectively.  The left-hand sides of the figures blow up 
the fixed-target and CERN ISR energy regime where the most data are
available while the right-hand sides
show the extrapolation of the cross sections to the collider regime.
Only a subset of the most recent fixed-target charm data are shown on the
right-hand side of Fig.~\ref{totcharm}.
The central value of the band is indicated by the solid curve while the upper
and lower edges of the band are given by the dashed curves.  The dotted curves
in Fig.~\ref{totcharm} are calculated with $\mu_F = \mu_R = 2m$ and $m=1.2$ GeV,
used in Ref.~\cite{Vogt:2001nh}.  Note that the charm uncertainty band broadens 
as the energy increases.  The lower edge of the charm band grows more slowly
with $\sqrt{S}$ above
RHIC energies while the upper edge is compatible with the reported total
cross sections at RHIC \cite{star,phenix}.

With $n_{\rm lf}$ light 
flavors and a fixed scale, the charm and bottom NLO total cross sections at 
$\sqrt{S} = 200$ GeV are 
\begin{eqnarray}
\sigma_{c \overline c}^{\rm NLO_{n_{\rm lf}}} & = & 
301^{+1000}_{-210} \, \, \mu{\rm b} \label{signlonlfcc} \, \, , \\ 
\sigma_{b \overline b}^{\rm NLO_{n_{\rm lf}}} & = & 
2.06^{+1.25}_{-0.81} \, \, \mu{\rm b} \label{signlonlfbb} \, \, .
\end{eqnarray} 
While the central values are only about 25\% and 10\% higher respectively
than the FONLL results in Eqs.~(\ref{sigccfonll}) and (\ref{sigbbfonll}), 
the uncertainty is considerably larger, especially for charm.
We now discuss the major sources of the theoretical uncertainty and
how the apparent discrepancy in the magnitude of the charm cross section
uncertainty in the RHIC results comes about.

\section{Comparison and discussion}
\label{comp-sec}

From the results in the previous two sections, it seems that the total
cross section is different depending on whether it is calculated from 
the integral over the inclusive $p_T$ distribution or from the total partonic
cross sections.  The difference seems especially large for charm production.
This is largely due to the way the strong coupling
constant is calculated and the low $x$, low scale behavior of the parton
densities.

In this section, we discuss these two contributions to the theoretical 
uncertainty and show that, if the total cross
section is calculated the same way, the two results are, in fact, equivalent,
as they should be.

\subsection{Strong coupling constant dependence}

The most trivial difference in the two calculations is that the $p_T$
distribution is calculated with a running scale proportional to $m_T$ while 
the total cross section is calculated with a fixed scale proportional to
$m$.  The charm quark uncertainty band is wider at low $p_T$, as shown in
Fig.~\ref{inclusive}, because $p_T \leq m$ and the calculation is
more sensitive to the lower scale in $\alpha_s$ since $m_T \sim m$ at low
$p_T$.  While it is more appropriate to use
the running scale to calculate inclusive distributions, the difference 
between a fixed and a running scale can be checked by fixing the scale
in the $p_T$ distributions.  The integral of the inclusive distribution
increases about 20\% for charm and about 10\% for bottom when a fixed scale
is used.  This difference is approximately large enough to account for the
difference in the central values of the total cross section.

One obviously important contribution to the uncertainty is the difference
in the number of flavors in the two calculations, especially for charm
since the fiducial range, $0.5 \leq \xi_R \leq 2$, is in a region where 
$\alpha_s$ is changing rapidly with $\mu_R$.  
Although increasing the number of light flavors involves more than just 
changing a parameter in the calculation of $\alpha_s$, we can get an estimate
of the importance of the value of $\alpha_s$ to the uncertainty in the total
cross section by looking at the dependence of $\alpha_s$ on the renormalization
scale.  When calculated with the
5 flavor QCD scale for CTEQ6M, $\Lambda_5 = 0.226$ MeV, and using a scheme
where $\alpha_s$ is continuous across mass thresholds, we have the
values shown in Table~\ref{tab:alphas}.
\begin{table}
\caption[]{The values of $\alpha_s$ for charm and bottom 
production at the given values of $\xi_R = \mu_R/m$.}
\begin{center}
\begin{tabular}{ccc}
$\xi_R$ & $n_{\rm lf}=3$, $m=1.5$ GeV & $n_{\rm lf}=4$, $m=4.75$ GeV \\ \hline
0.5 & 0.6688 & 0.2822 \\
1   & 0.3527 & 0.2166 \\
2   & 0.2547 & 0.1804 \\ \hline
\end{tabular}
\end{center}
\label{tab:alphas}
\end{table}
It is clear, based on these values alone, that the charm uncertainty is larger
than that for bottom since $\alpha_s(\xi_R = 0.5)/\alpha_s(\xi_R = 2) =
2.63$ for charm and 1.56 for bottom.  The real difference in coupling strength 
between the two heavy quarks is even larger since the leading order cross
section is proportional to $\alpha_s^2$ while the next-order contribution is
proportional to $\alpha_s^3$.

Using $n_{\rm lf}+1$ in the FONLL and NLO calculations of the inclusive 
distributions in Section~\ref{fonll-sec} 
reduces the uncertainty.  When the total cross sections in 
Eqs.~(\ref{sigccfonll}) and (\ref{sigccfonll-nlo}) are instead
calculated with $n_{\rm lf}$, the uncertainty is increased so that the 
upper and lower limits of the charm uncertainty are in agreement with 
Eq.~(\ref{signlonlfcc}) \cite{matteopriv}.  Thus whether charm is treated as
a heavy ($n_{\rm lf}$) or an active ($n_{\rm lf} + 1$) flavor in the calculation
turns out to be one of the most important influences on the limits of the charm
uncertainty.

\subsection{Parton density dependence}  

Next, we discuss the influence of the parton densities on the theoretical
uncertainty.  Since $m$ is the only perturbative scale, the total cross section
calculations in Section~\ref{nlo-sec} are more
sensitive to the low $x$ and low $\mu$ behavior of the parton densities.
Probing the full fiducial range of the uncertainty band is problematic for
charm production since $\xi_F = 0.5$ is below the minimum scale of the 
CTEQ6M parton densities, $\mu_0^{\rm CTEQ6M} = 1.3$ GeV.  Thus, for this scale, 
backward evolution of
the parton densities is required.

\begin{figure}[htbp]
\setlength{\epsfxsize=0.95\textwidth}
\setlength{\epsfysize=0.3\textheight}
\centerline{\epsffile{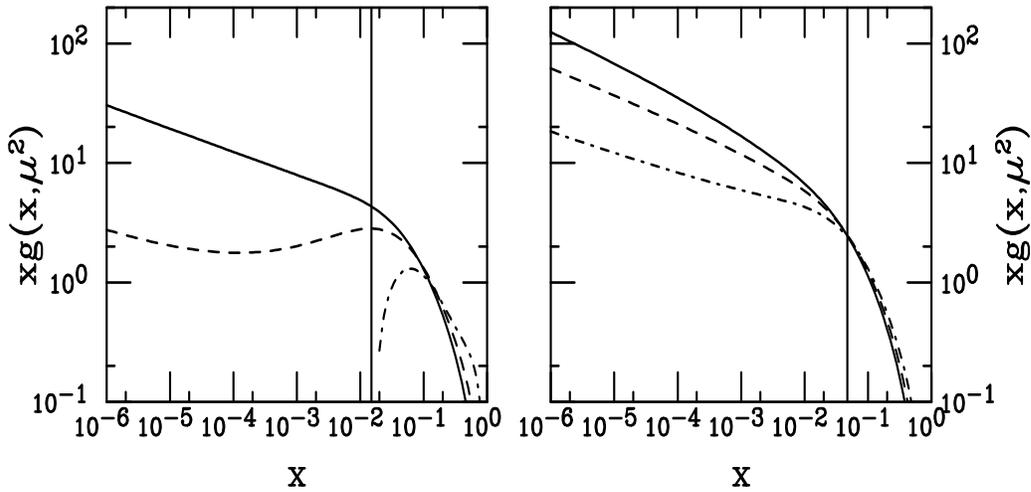}}
\caption[]{The CTEQ6M parton densities as a function of $x$ for $\xi = 0.5$ 
(dot-dashed), $\xi = 1$ (dashed) and $\xi = 2$ (solid) for $m = 1.5$ GeV
(left-hand side) and 4.75 GeV (right-hand side).  The vertical line is
the value $x = 2m/\sqrt{S}$ in $\sqrt{S}=200$ GeV $pp$ collisions at RHIC.
}
\label{cteqpdf}
\end{figure}

The CTEQ6M (NLO, $\overline{\rm MS}$ scheme) gluon distributions in the 
fiducial region of the factorization scale, $0.5 \leq \xi_F \leq 2$, are 
shown in Fig.~\ref{cteqpdf}.  The behavior of the gluon distributions
for charm (left) and bottom (right) are quite different.  Since the range
$0.5m \leq \mu_F \leq 2m$ for bottom quarks lies well above $\mu_0^{\rm CTEQ6M}$,
the scale dependence of the gluon density is typical.  The gluon density
increases with decreasing $x$.  The highest low $x$ gluon density is at 
the largest scale.  For $x$ values larger than that of central rapidity at
RHIC, to the right of the vertical lines in Fig.~\ref{cteqpdf}, the 
gluon densities are rather similar although the density is larger at the lower
scale.  This $x$ dependence is quite 
typical for large, perturbative factorization scales and demonstrates why the
bottom quark cross section is well behaved as a function of $\sqrt{S}$.

The gluon distributions with $\xi_F=2$ for charm and $\xi_F = 0.5$ for bottom 
are similar because $\mu_F = 3$ GeV and
2.375 GeV respectively. Thus at the highest $\xi_F$ for charm, the low $x$
gluon density is well behaved.  However, the behavior at lower
scales is quite different, especially for $x < 10^{-2}$.  When $\xi_F = 1$,
the dashed curve on the left-hand side of Fig.~\ref{cteqpdf} is no longer
increasing with decreasing $x$ but, instead, is almost flat for $x < 10^{-2}$
with a slight dip in the middle.
Lower $x$ values are not shown for $\xi_F = 0.5$ because the
backwards evolution gives $xg(x,\xi_F = 1) = 0$, accounting for the high
$\sqrt{S}$ behavior of the lower bound on the uncertainty band.
The low $x$, low $\mu_F$ behavior of 
the gluon density depends strongly on how the group performing the global
analysis chooses to extrapolate to unmeasured regions.  All that is 
required is minimization of
the global $\chi^2$ and momentum conservation.  

\subsection{Scale dependence}

Finally, we describe the scale dependence of the charm and bottom cross
sections in some detail.

\subsubsection{Bottom}

We first focus on bottom production because the factorization scale is larger
than $\mu_0^{\rm CTEQ6M}$ in the entire fiducial region.  In general,
when the factorization scale in Eq.~(\ref{sigpp})
is sufficiently high, the lowest scales give the highest cross sections
because the evolution to higher scales reduces the gluon density at 
$x > 10^{-2}$.  The largest cross sections in the fixed-target regime are 
thus obtained with the combinations $(\xi_R,\xi_F)=(0.5,0.5$),
(0.5,1) and (1,0.5).  At relatively low $\sqrt{S}$, the slightly higher
value of $\alpha_s(\xi_R=0.5)$ compensates for the lower gluon
density with $\xi_F=1$ at large $x$ so that the cross section with 
$(\xi_R,\xi_F) = (0.5,1)$ is higher than that with (1,0.5).  As the energy
increases and lower $x$ values are probed, the lower small $x$ gluon density
with $\xi_F = 0.5$ can no longer overcome the difference between $\alpha_s(\xi_R
=1)$ and $\alpha_s(\xi_R=0.5)$ and, at LHC energies, the cross section with
$(\xi_R,\xi_F)=(1,0.5)$ is smaller.  Indeed, the lower factorization scale
with $(\xi_R,\xi_F)=(0.5,0.5)$ is not enough to keep the cross section with this
parameter set larger than that with (1,0.5) or, for that matter, those with
either or both $\xi_R, \xi_F = 2$ at sufficiently low $x$.

A similar effect occurs for $(\xi_R,\xi_F)=(2,2)$, (2,1) and (1,2) except that
now, at fixed-target energies, the set $(\xi_R,\xi_F)=(2,2)$ 
gives the lowest cross section of the three pairs. In this case, 
$(\xi_R,\xi_F)=(1,2)$ gives the highest cross section of the
three sets since the gluon density with $\xi_F= 1$ is higher at large $x$ (lower
$\sqrt{S}$) and $\alpha_s(\xi_R=1)$ is larger, compensating for the
slightly lower gluon density at large $x$.  
At collider energies, $(\xi_R,\xi_F)=(1,2)$ still gives the largest 
cross section of these three parameter sets since the evolution at 
low $x$ (large $\sqrt{S}$) is the dominant behavior. At large $\sqrt{S}$, 
the cross section with $(\xi_R,\xi_F) = (2,1)$ drops below those calculated with
the other two sets.

These subtle changes in which $(\xi_R,\xi_F)$ set dominates the
upper and lower limits of the bottom quark total cross section uncertainty 
band as a function of $\sqrt{S}$ do not significantly broaden the uncertainty
band, even at the highest energies because $\mu_F > \mu_0^{\rm CTEQ6M}$.  

The scale choice in the parton densities affects the dominance of a 
particular parameter set ($\xi_R,\xi_F)$ in the $p_T$ distributions to
a lesser extent because at $p_T > m$ the scales are all large and perturbative.
Different parameter sets dominate the $p_T$ distribution because, at RHIC, 
high $p_T$ probes the large $x$ range of the gluon distribution while for
$p_T \rightarrow 0$, $x$ is relatively small.  At $p_T \rightarrow 0$, 
the upper and lower edges of the band are thus 
determined by $(\xi_R,\xi_F) = (0.5,1)$ and (1,0.5) respectively,
as is also the case for the total cross sections.  However, as $p_T$ increases,
the upper and lower edges of the band are defined by $(\xi_R,\xi_F) = (0.5,0.5)$
and (2,2) respectively.  Increasing $p_T$ has the same effect as moving
to smaller $\sqrt{S}$: both probe larger $x$ where the gluon distribution
with $\xi_F = 0.5$ is higher than that with $\xi_F = 2$, as is obvious from
the right-hand side of Fig.~\ref{cteqpdf}.

Even though the scale dependence of bottom production is not negligible,
as we have seen, it is not strong in the defined fiducial range.
The difference in the $b \overline b$ cross sections in Eqs.~(\ref{sigbbfonll})
and (\ref{signlonlfbb}) can be almost entirely attributed to the change from 
the running scale in Eq.~(\ref{siginclusive}) and the fixed scale in 
Eq.~(\ref{sigpp}).  Thus the bottom production cross section is rather 
well under control.  

\subsubsection{Charm}

However, the scale dependence of the total charm cross section on
$\sqrt{S}$ is another story due to the behavior of the CTEQ6M gluon 
distribution at charm quark scales.
Since $m \sim \mu_0^{\rm CTEQ6M}$, using the full fiducial region to
estimate the theoretical uncertainty on the total charm cross section
problematic.  The smaller charm mass exaggerates the factorization
scale dependence of the total
cross section described above for bottom production.

Thus the charm quark uncertainty band on the total cross section,
Fig.~\ref{totcharm}, spans an order of magnitude 
at fixed-target energies, increasing to the value given in 
Eq.~(\ref{signlonlfcc}) for $n_{\rm lf}$ at $\sqrt{S} = 200$ GeV.  The
low scale behavior for $(\xi_R,\xi_F)=(0.5,1)$ and (1,0.5) defines the upper
and lower edges respectively
of the uncertainty band at collider energies.
Indeed, for the total cross section calculated with $n_{\rm lf}$ light quark
flavors, the STAR point \cite{star} is compatible with the upper limit of the
band although the inclusive $p_T$ data lies above the FONLL calculation
with $n_{\rm lf}+1$ light flavors \cite{diff}.  However, we stress that this
apparent agreement of the STAR result with the total cross section does not
mean that the discrepancy between the high $p_T$ STAR results and the FONLL
prediction can be ignored.  At high $p_T$, the FONLL calculation is more
reliable since here charm is correctly treated as an active flavor, with
$n_{\rm lf} + 1$, and light quark effects are resummed, improving the prediction
at finite $p_T$.

The charm band grows broader with increasing $\sqrt{S}$, corresponding 
to decreasing $x$.
At 10 TeV, the width of the uncertainty band has increased to almost two orders
of magnitude.  Thus, without a better handle on the gluon density at low $x$
and low scales, one may question whether such a large uncertainty is 
meaningful.  It may also be questionable whether the lowest scales, 
$\xi_R$, $\xi_F = 0.5$ should be included in the calculation of the charm
uncertainty, especially when $\mu_F < \mu_0^{\rm CTEQ6M}$ for three light flavors.

The full fixed-target data set also exhibits a large uncertainty
due to the method of extrapolation used, the assumed branching ratios and the
$A$ dependence, as shown on the left-hand side of Fig.~\ref{totcharm}.
However, if only the most recent data are used, the uncertainty in the 
data seems to be reduced.
As an alternative, one may try to `fit' the mass and scale parameters to these
data \cite{Vogt:2001nh} for $\mu > m$.  
The dotted curves in Fig.~\ref{totcharm} show the energy dependence of 
one such attempt with $m=1.2$ GeV, $(\xi_R,\xi_F)=(2,2)$.  The calculated
cross section lies just above the central value of the band and, 
although the quark mass is smaller than the assumed central mass value, 
the larger value of $\xi_F$
guarantees a more regular $\sqrt{S}$ dependence than that obtained with smaller
values, as shown in Fig.~\ref{cteqpdf}.

\section{Conclusions}

We have shown that when the total cross section is calculated 
with the same parameter sets and
the same number of light quark flavors, a consistent result is obtained
by both integrating over an inclusive distribution and starting from the total
partonic cross section, as should be expected.  However, the charm
results are extremely sensitive to the number of flavors, the scale choice
and the parton densities.  One of the biggest sources of
uncertainty in the total charm cross section at collider energies
is the behavior of the
gluon density at low $x$ and low scale, as yet not well determined.  Until
it is further under control, better limits on the charm quark total cross
section will be difficult to set.  A complete NNLO evaluation of the total
cross section may reduce the scale dependence but will still be subject to
the same types of uncertainties.

It is thus not clear which estimate of the total charm cross section
uncertainty, Eqs.~(\ref{sigccfonll}) and (\ref{sigccfonll-nlo}) or 
Eq.~(\ref{signlonlfcc}), is more reliable.  If the low $p_T$ region 
is ignored and the
heavy quark may be considered an active flavor then the appropriate number
of flavors is $n_{\rm lf}+1$ rather than $n_{\rm lf}$ and the smaller error 
band used to compare the RHIC $p_T$ distributions \cite{diff}
is more reasonable.  However, when heavy flavor production is measured over
the full $p_T$ range, down to $p_T \sim 0$, then three light flavors should
likely be used for charm, resulting in the larger
uncertainty.  Unfortunately, in this case, the uncertainty is driven by scales
lower than the initial scale of the parton density, further complicating the
interpretation of the limits on the uncertainty band.  Thus, rather than
arbitrarily choosing one result over another, we prefer to stress
that there is little predictive power in the charm production uncertainty.

\section*{Acknowledgements}

I am very pleased to contribute to this volume in honor of Prof. Zimanyi.
I would like to thank him for introducing me to Budapest and Hungarian culture.
From my very first trip to Budapest, Joszo and Magda were very kind to me,
making sure I saw something besides the interior of the laboratory.  Thanks to
Joszo and his very active group, I have enjoyed many visits over the years
and look forward to more in the future. 

I would like to thank M. Cacciari, G. Odyniec and T. Ullrich for discussions. 
This work was performed under the auspices of the U.S. Department of Energy by 
University of California, Lawrence Livermore National Laboratory under 
Contract W-7405-Eng-48 and was also supported in part by the National Science
Foundation Grant NSF PHY-0555660.

\end{document}